\documentstyle[aps,pre,epsf,graphicx,amsmath,array]{revtex}
\begin{document}

\title{Statistics of the contact network in frictional and
  frictionless granular packings}
 
\author{Leonardo E.~Silbert,\footnote{ Present address: James Franck
    Institute, University of Chicago, Chicago, Illinois 60637} Gary S.~Grest,
  and James W. Landry}
\address{Sandia National Laboratories, Albuquerque, New Mexico 87185}
 
\maketitle 

\begin{abstract}
  Simulated granular packings with different particle friction
  coefficient $\mu$ are examined. The distribution of the
  particle-particle and particle-wall normal and tangential contact
  forces $P(f)$ are computed and compared with existing experimental
  data. Here $f\equiv F/\bar{F}$ is the contact force $F$ normalized
  by the average value $\bar{F}$. $P(f)$ exhibits exponential-like
  decay at large forces, a plateau/peak near $f=1$, with additional
  features at forces smaller than the average that depend on $\mu$.
  Computations of the force-force spatial distribution function and
  the contact point radial distribution function indicate that
  correlations between forces are only weakly dependent on friction
  and decay rapidly beyond approximately three particle diameters.
  Distributions of the particle-particle contact angles show that the
  contact network is not isotropic and only weakly dependent on
  friction. High force-bearing structures, or force chains, do not
  play a dominant role in these three dimensional, unloaded packings.
\end{abstract}


\section{Introduction}
The properties of granular materials, even static sandpiles, continue
to perplex engineers and physicists alike \cite{nagel1}. Processing of
granular materials play a central role in the pharmaceuticals industry
as well as engineering communities, such as ceramic component design.
For example, one may wish to evenly distribute the ingredients in a
tablet or pill, or reduce the likelihood of component failure.
However as yet there is no clear indication of how the individual
particle properties determine the final state of the system.

Although packings of frictionless, monodisperse, cohesionless, hard
spheres have been well studied \cite{combo1}, little is known about
the effect of including particle friction.  Recent discrete element
simulations of granular materials, where packings were generated for
particles for different static coefficients of friction $\mu$ between
individual particle pairs, showed that the local particle coordination
of the packing varied strongly as a function of friction \cite{leo9}.
From a different perspective, experimental studies of static granular
assemblies have shown many interesting facets of the stress state of
these systems.  One method of analysis appears to dominate in
describing the statistics of granular packings: computations of the
distributions of normal forces are {\it de rigeur}.

Experimental studies on granular packings use a variety of techniques
to measure the distribution of normal contact forces $P(f)$ between
particles and container walls
\cite{mueth1,coppersmith1,nagel5,nagel6,lovoll1,charmet1}, where
$f\equiv F/\bar{F}$ - all measured normal forces $F$ are normalized
with respect to the average force $\bar{F}$. The Chicago group
\cite{mueth1,coppersmith1,nagel5,nagel6} utilized carbon paper to
measure $P(f)$ at the base and sides of a cylindrical container packed
with glass spheres with a normal load applied at the top of the
packing. Forces several times the average force were observed, with
resolution down to the weight of a few particles. Blair et
al.~\cite{nagel5} measured $P(f)$ for amorphous and ordered granular
packings for particles with different values of $\mu$, which varied by
a factor of approximately three. These experiments demonstrated that
$P(f)$ is indiscriminate towards the effects of particle friction and
structure of the packing, and the general form of $P(f)$ remained
robust within the resolution of the experiment.

In a different experimental set up, L$\phi$voll et al.~\cite{lovoll1}
used a pressure transducer device to measure $P(f)$ at the bottom of
an unloaded granular packing under its own weight, on a fixed
substrate of particles glued to the supporting base. This experiment
was able to resolve forces down to the weight of a few grain masses
and showed that the spatial distribution of contact forces were
correlated over a few particle diameters. Using a novel modification
of the carbon paper technique, Tsoungui et al.~\cite{charmet1}
actually measured $P(f)$ inside the bulk of a 2D packing. Despite the
poorer statistics of this study, the results agreed well with Blair et
al.~\cite{nagel5} and L$\phi$voll et al.~\cite{lovoll1}. Experimental
studies on static granular packings show that $P(f)$ exhibits several
generic features; an approximately exponential tail at large $f$ and a
plateau or peak near $f\approx 1$.

Computational studies, such as contact dynamics or molecular dynamics,
of compressed packings provide good comparison with the experimental
data \cite{radjai1,radjai2}. However, as yet, there has been no
systematic study of the effects of particle friction on the force
distributions within a granular assembly. Here we show how the effects
of friction change the behaviour of $P(f)$ in the small force region
but only weakly affect the large-$f$ region. We show that the
local contact geometry of the packing is not isotropic and only weakly
influenced by friction. We also discuss aspects of the force network
whereby high force-bearing structures, or force chains, do not seem to
be a dominant feature of these unloaded packings.

We computed $P(f)$ in the bulk of various packings (which is presently
inaccessible in 3D experiments) that had settled onto either a rough
bed or a planar base. We compared these results with $P(f)$ for
particles in contact with the flat base (similar to experiment) of a
periodic packing and with the $P(f)$ generated at the side walls of a
cylindrical packing. We resolve the components of the contact force
that are normal ($n$) and tangential ($t$) to the line of centres
between two particles in contact.

In the next section we briefly describe the model, though a more
thorough description of the technique is available elsewhere
\cite{leo9,leo7}. In section \ref{results}, we present results for the
force distributions, force correlations, and the contact geometry. We
also discuss some aspects of the force network with respect to a force
cut-off scheme, highlighting some pros and cons of this method. In
section \ref{summary} we summarize and conclude this work.

\section{Model and Method}
We performed three dimensional (3D) molecular dynamics simulations
with $N$ monodisperse, cohesionless, inelastic spheres that interact
only on contact via a Hooke (linear) spring or a Hertz contact law and
static friction \cite{cundall1,walton2}. Contacting particles $i$ and
$j$ positioned at ${\bf r}_{i}$ and ${\bf r}_{j}$ experience a
relative normal compression $\delta=|{\bf r}_{ij}-d|$, where ${\bf
  r}_{ij}={\bf r}_{i}-{\bf r}_{j}$, which results in a force ${\bf
  F}_{ij}={\bf F}_{n}+{\bf F}_{t}$. The normal and tangential contact
forces are given by,
\begin{equation}
{\bf F}_{n}=f(\delta/d)\left(k_{n}\delta 
{\bf n}_{ij}-\frac{m}{2}\gamma_{n}{\bf v}_{n}\right), 
\label{equation2}
\end{equation}
\begin{equation}
{\bf F}_{t}=f(\delta/d)\left(-k_{t}\Delta{\bf s}_{t}
-\frac{m}{2}\gamma_{t}{\bf v}_{t}\right),
\label{equation3}
\end{equation}
where ${\bf n}_{ij}={\bf r}_{ij}/r_{ij}$, with $r_{ij}=|{\bf
  r}_{ij}|$, ${\bf v}_{n}$ and ${\bf v}_{t}$ are the normal and
tangential components of the relative surface velocity, and $k_{n,t}$
and $\gamma_{n,t}$ are elastic and viscoelastic constants
respectively. $f(x)=1$ for Hookean springs and $f(x)=\sqrt{x}$ for
Hertzian contacts. $\Delta{\bf s}_{t}$ is the elastic tangential
displacement between spheres, obtained by integrating surface relative
velocities during elastic deformation of the contact. The magnitude of
$\Delta{\bf s}_{t}$ is truncated as necessary to satisfy a local
Coulomb yield criterion $F_{t} \leq \mu F_{n}$, where $F_t \equiv
|{\bf F}_{t}|$ and $F_{n} \equiv |{\bf F}_{n}|$, and $\mu$ is the
particle-particle friction coefficient. For the present simulations we
set $k_{n}=2\cdot10^{5}mg/d$, $k_{t}=\frac{2}{7}k_{n}$,
$\gamma_{n}=50\sqrt{g/d}$. For Hookean springs we set $\gamma_{t}=0$
while for Hertzian springs,  $\gamma_t=\gamma_n$. For Hookean
springs the coefficient of restitution $\epsilon_{n,t}$, is related to
$\gamma_{n,t}$ through, \[ \epsilon_{n,t} = \exp(-\gamma_{n,t}t_{\text
  col}/2), \] where the collision time $t_{\text col}$ is determined
by the contact frequency between two particles. For the parameters
chosen, $\epsilon_{n} = 0.88$ for Hookean springs. For Hertzian
contacts $\epsilon$ is velocity dependent\cite{wolf1}. We chose a
time-step $\delta t=10^{-4}\tau$, where $\tau=\sqrt{d/g}$ and $g$ is
the gravitational acceleration.

Amorphous packings (with packing fraction $\phi\approx0.60$) were
generated by allowing an initially dilute system to settle under
gravity acting in the vertical direction. Particles settled onto a
bottom wall that was either a planar base or a bumpy bed of particles
frozen into a close packed random configuration. This process was run
until the kinetic energy of the system was much smaller than the
potential energy \cite{leo9}. The base had the same frictional and
elastic properties as the particles being poured. 

Most of our results are for packings that are spatially periodic in
the horizontal plane, i.e. we ignored the effects of sidewalls.
Because of this, the pressure in a packing does not saturate with
depth.  Therefore, to make a direct comparison with experiment, our
definition of the average force $f_{n,t} \equiv F_{n,t} /
\bar{F}(z)_{n,t}$, was normalized by $\bar{F}(z)_{n,t}$, the average
contact force at a depth $z$ in the packing. The generation of these
packings is fully discussed in Ref.~\cite{leo9}. We also compared
results for packings poured into a cylindrical container with `flat'
walls and the same properties as the particles. In this case, there is
no need for depth-average normalisation, as the walls carry a
significant fraction of the weight of the system \cite{landry1}.
Results for the periodic packings with depth-average normalisation are
consistent with the cylindrically confined packings and experiment. As
a consequence, the depth-average normalisation proves to be the
correct method for dealing with periodic packings.

\section{Results}
\label{results}
\subsection{Force Distributions}
Force distributions in all granular packings exhibit several general
features. Measurements of the distribution of normal contact forces
$P(f_{n})$, for granular packings that are either free-standing under
the influence of gravity \cite{lovoll1} (as we simulate here),
confined packings that have been loaded (as in experiments)
\cite{mueth1,nagel5,charmet1}, or axially compressed systems (as in
previous simulation studies and experiment)
\cite{radjai1,radjai3,thornton1}, as well as a lattice model
\cite{coppersmith1}, all purport exponential tails in $P(f_{n})$ at
large forces (typically for $f_{n}>1$). Mueth et al.~\cite{mueth1}
used an empirical fit to their experimental data of the form,
\begin{equation}
P(f_n) = a(1-be^{-f_n^{2}})e^{-\beta f_n},
\label{equation1}
\end{equation}
and found $a=3.0$, $b=0.75$, and $\beta = 1.5\pm 0.1$ for loaded glass
spheres confined in a cylindrical container.

\begin{figure}[t]
\begin{center}
\includegraphics[width=6cm]{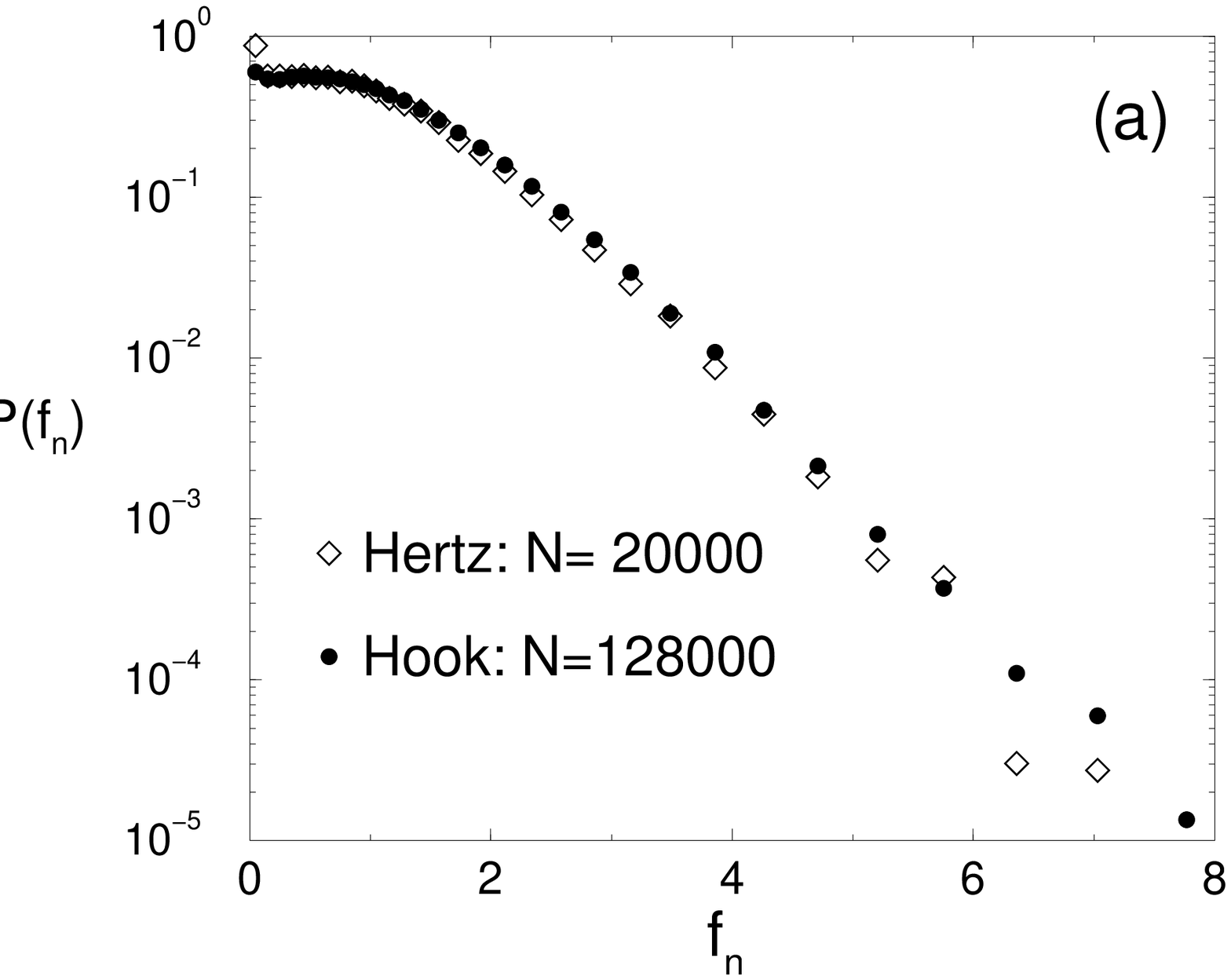}
\hfil\includegraphics[width=6cm]{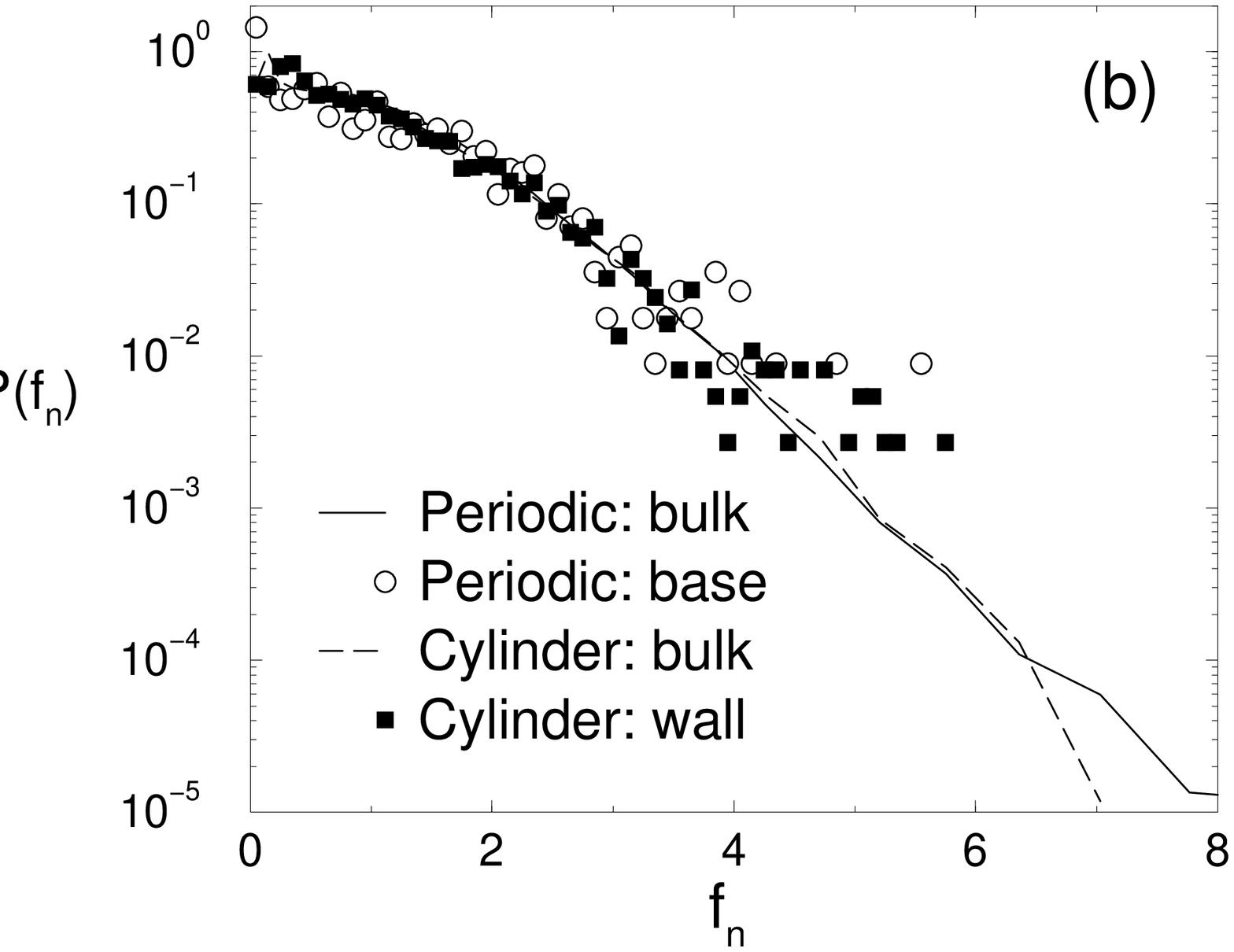}
\caption{Distribution of {\it normal} contact forces $P(f_{n})$ for packings
  of $N$ monodisperse spheres of diameter $d$ and particle friction
  coefficient $\mu=0.5$. (a) Comparison between a spatially periodic
  Hertzian packing with a square base of dimensions $A=20d\times 20d$
  and a Hookean packing with $A=40d\times 40d$. System sizes are
  indicated in legend. (b) Comparison between two Hookean packings,
  one a spatially periodic system with $N=128000$ and $A=20d\times
  20d$, the other a confined, cylindrical packing of diameter $D=20d$
  and $N=50000$.}
\label{figure1}
\end{center}
\end{figure}

In Fig.~\ref{figure1} we show our computations of the force
distributions for the normal contact force $f_{n}$ for different
systems. In Fig.~\ref{figure1}(a) we see that the form of $P(f_{n})$
is the same for both Hookean or Hertzian contact force laws. Varying
the system size has no effect (other than improving the statistics of
the data). Similarly, in Fig.~\ref{figure1}(b) computations of $P(f_n)$
in the bulk of a periodic or confined system, at the base of the
periodic system, or at the sidewalls of the cylinder are
indistinguishable. Recent 2D simulations have shown that $P(f_n)$ at the
base can depend on the properties and geometry of the base
\cite{snoeijer1}. Computations of $P(f_{n})$ for those particles in
contact with the flat base and at the side walls also show the generic
form seen in the other data although the statistics here are poor due
to the number of contacts in the plane ($\approx 10^{4}$) compared
with the number of particle-particle contacts in the bulk ($\approx
10^{5} - 10^{6}$).

\begin{figure}[t]
\begin{center}
  \includegraphics[width=6cm]{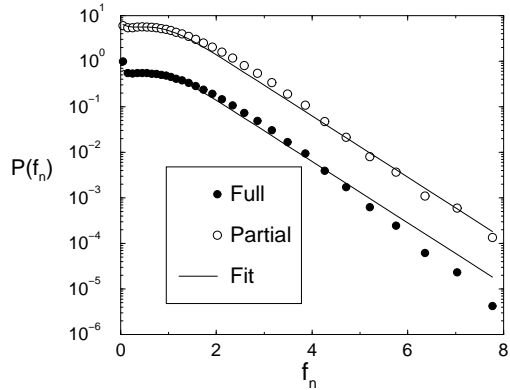}
\caption{Distribution of {\it normal} contact forces $P(f_{n})$ for Hookean
  packings of $N=128000$ monodisperse spheres and $\mu=0.5$, on a flat
  base of dimensions $40d\times 40d$. The {\it full} $P(f_n)$ (solid
  circles) includes normal forces for all contacting particles and we
  fit to Eq.~\ref{equation1} (solid line) using $a=2.55$, $b=0.65$,
  and $\beta = 1.32$. For the {\it partial} $P(f_n)$ (open circles) we
  have excluded all forces less than the weight of one grain and
  renormalized $f$, finding a better fit to Eq.~\ref{equation1} with
  $a=3.1$, $b=0.78$, and $\beta = 1.55$. We have arbitrarily shifted
  the curve for the  {\it partial} $P(f_n)$ for clarity.}
\label{figure2}
\end{center}
\end{figure}

To compare with existing experimental data, we fit Eq.~\ref{equation1}
to our data for the largest system. We show this comparison in
Fig.~\ref{figure2}. The $P(f_{n})$ computed over all contact forces is
denoted by the solid circles in Fig.~\ref{figure2} with best fit
parameters $a=2.55$, $b=0.65$, and $\beta = 1.35$, agreeing well with
Eq.~\ref{equation1} up to $f\approx 2$, but falling off more quickly
than Eq.~\ref{equation1} for large $f$. We find a better fit to
Eq.~\ref{equation1} if we exclude the data for $F_{n}<mg$,
i.e.~ignoring all data in the limit $f\rightarrow 0$, essentially
mimicking the finite resolution in experiment. This alters the average
value such that our original data set has now been `squeezed'
together. We denote this data as the {\it partial} set in
Fig.~\ref{figure2}. The fit to Eq.~\ref{equation1} with $a=3.1$,
$b=0.78$, and $\beta = 1.55$, is much better than when data for small
forces is included. Our simulation data is in good qualitative and
quantitative agreement with previous experimental results
\cite{mueth1} and similar to Radjai et al.~\cite{radjai1}.

The empirical fit of Eq.~\ref{equation1} is poor for the total bulk
$P(f_{n})$ over a large range of the data and we only achieve
agreement by renormalising our data, using the partial data set in
Fig.~\ref{figure2}. We also note that on closer inspection of existing
simulation and experimental data, whether the tails of $P(f_{n})$ are
truly exponential or not is questionable and may be an indication of
the averaging technique used in computational studies \cite{ohern2}.

\begin{figure}[t]
\begin{center}
  \includegraphics[width=6cm]{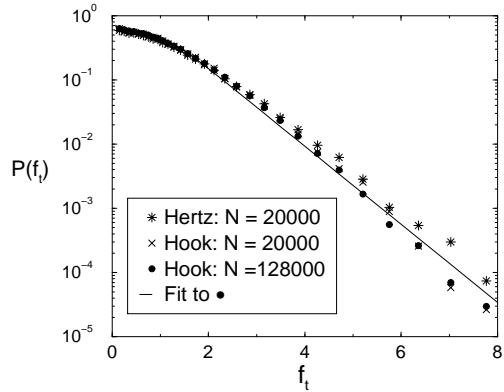}
\caption{Distribution of {\it tangential} contact forces $P(f_{t})$ for
  packings of $N$ monodisperse spheres of diameter $d$, with particle
  friction coefficient $\mu=0.5$. System size is shown in legend. The
  line is fit to Eq.~\ref{equation1} for the largest system.}
\label{figure3}
\end{center}
\end{figure}

The distribution of tangential forces $P(f_{t})$ is shown in
Fig.~\ref{figure3}. In comparison with the normal forces, $P(f_{t})$
decays more slowly than $P(f_{n})$. Fitting Eq.~\ref{equation1} to the
bulk data for the largest system ($N=128000$), we find good agreement
with $a=2.5$, $b=0.7$, and $\beta=1.4$.

\begin{figure}[t]
\begin{center}
  \includegraphics[width=6cm]{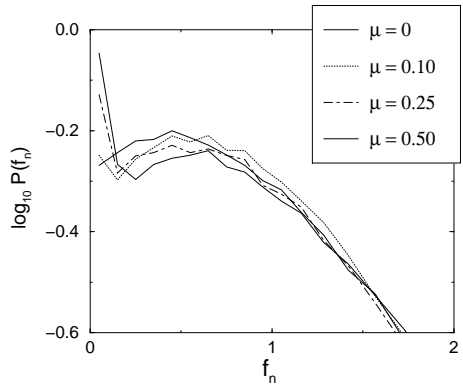}
\caption{$P(f_{n})$ at small forces for packings with different particle
  friction coefficient $\mu$. Frictionless packings ($\mu=0$) exhibit
  a well-defined peak in $P(f_{n})$ near $f_{n}=1$, whereas even for
  low frictional packings, an upturn appears in $P(f_{n})$ at very
  small forces. The amplitude of this upturn increases with increasing
  friction coefficient and the position of the peak also shifts to 
  larger $f_n$. Results are for Hookean packings with periodic boundary
  conditions in the horizontal plane, for $N=20000$ on a rough,
  particle base with $A=20d\times 20d$.}
\label{figure4}
\end{center}
\end{figure}

While there is clearly some agreement on the behaviour of $P(f_n)$ for
large $f_n$, the characteristic nature of the small force region of
$P(f_{n})$ remains in dispute. Experimental data show that $P(f_{n})$
approaches a finite value as $f_{n}\rightarrow 0$. However, some
numerical works have suggested that $P(f_{n})\rightarrow 0$ for small
$f_{n}$ \cite{coppersmith2}. In Fig.~\ref{figure4} we show the small
force region of $P(f_{n})$ for packings identically prepared but with
different particle friction coefficients. We do not show the full
$P(f_n)$ curve as friction only weakly influences the behaviour of the
large-$f$ region. However, our large system size data suggests that
the exponential tail becomes slightly steeper with decreasing
friction, i.e.~$\beta$ increases as $\mu$ decreases. The defining
feature of these packings is that for purely frictionless systems,
$P(f_{n})$ shows a well-defined peak at small forces, while for
$\mu>0$, $P(f_{n})$ has an upturn at very small forces. The amplitude
of this upturn increases with increasing friction coefficient.

The Chicago group \cite{nagel5} studied different particle packings
where $\mu$ varied by a factor of approximately three. Within the
resolution of their experiment they did not find any systematic trend
with friction. Because of the higher resolution in simulation, the
following comments are relevant to such studies: the fraction of
particle-particle contacts, or {\it bonds}, experiencing small forces
increases with increasing $\mu$ even though the total number of
contacts decreases with increasing $\mu$ \cite{leo9}. Further study
shows that the fraction of particles that are close to the Coulomb
yield criterion $F_{t}\approx\mu F_{n}$, i.e. those particle pairs
that are most likely to undergo local plastic rearrangement, increase
as $\mu \rightarrow 0$.  Indeed, we have previously reported
\cite{leo9} that frictionless packings are always isostatic, whereas
frictional packings are hyperstatic and this may be related to the
behaviour of $P(f_n)$ at small $f_n$.

\begin{figure}[t]
\begin{center}
  \includegraphics[width=6cm]{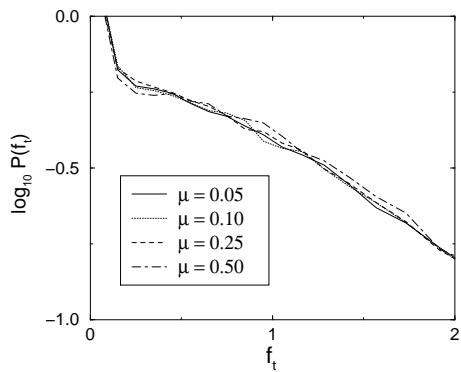}
\caption{$P(f_{t})$ at small forces for packings with different particle
  friction coefficient $\mu$. Results are for a Hookean packing with
  periodic boundary conditions for $N=20000$ and $A=20d\times 20d$.}
\label{figure5}
\end{center}
\end{figure}

For completeness we show the corresponding distributions $P(f_{t})$,
for the tangential forces in Fig.~\ref{figure5}. In this case, we do
not find any significant systematic trend with $\mu$. The role of
$\mu$ in the determination of $P(f_{n,t})$, is subtle. In frictionless
packings, $P(f_{n})$ does not show an upturn at small $f_{n}$,
therefore the generation of this upturn in frictional packings comes
from the very presence of the frictional forces $f_{t}$, which
influence the nature of particle contacts such that $P(f_{n})$ itself
observes an upturn at small forces.

\subsection{Force Correlations}
The spatial force-force correlation function ${\mathcal F}(r)$
measures spatial correlations between forces separated by a distance
$r$. We use the same definition as in Refs.~\cite{mueth1,lovoll1},
\begin{equation}
{\mathcal F}(r)\equiv\frac{\sum_{i}\sum_{j>i}\delta(|{\bf r}_{ij}|-r)f_{i}f_{j}}
{\sum_{i}\sum_{j>i}\delta (|{\bf r}_{ij}|-r)}~,
\label{equation4}
\end{equation}
where ${\bf r}_{ij}$ is the distance between particle contacts $i$ and
$j$, and $f_{i}$ is the normalized contact force acting at contact
$i$. In experiment, spatial force correlations can, at present, only
be measured at container walls: the points of force contact coincide
with particle contacts at the container surface lying in a 2D plane.
The minimum separation between measurements in experiment is
coincident with the particle size, $r_{\text min}\approx d$. In a 3D
packing, contact forces transmitted at the points of particle-particle
contacts are only restricted by excluded volume effects. For
monodisperse spheres in 3D the minimum separation, $r_{\text
  min}\approx \frac{d}{2}$. A locally four-particle pyramid
configuration would give this minimum separation. Mueth et
al.~\cite{mueth1} found no evidence for spatial correlations between
the contact forces within the resolution of their measurements.
L$\phi$voll et al.~\cite{lovoll1}, using a different measuring
technique, resolved their force data showing weak force correlations
at the base of their packings which extend out to approximately five
particle diameters \cite{lovoll1}. This may only come about from the
induced order of the sample at the container wall.

\begin{figure}[t]
\begin{center}
  \includegraphics[width=6cm]{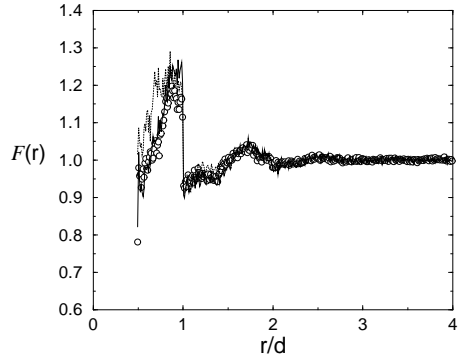} \bigskip
\caption{Spatial force-force correlation function ${\mathcal F}(r)$ for normal
  contact forces as defined in Eq.~\ref{equation4} computed within the
  bulk of a frictionless (circles) and frictional ($\mu=0.5$ -- solid
  line) packing. The dotted line shows the corresponding correlation
  for the tangential forces when $\mu=0.5$.}
\label{figure6}
\end{center}
\end{figure}

Because of the restricted geometry of experimental measurements, we
found it instructive to compare our computations of ${\mathcal F}(r)$
for the normal contact forces within the bulk of amorphous packings,
and see how these might depend on $\mu$. For comparison we also
computed the correlation function between tangential contact forces
for $\mu=0.5$. In Fig.~\ref{figure6} we present the spatial force
correlation function for a frictionless packing ($\mu = 0$) and a
frictional packing ($\mu=0.50$). Within the bulk of the packing,
forces are correlated, but only over short distances, extending to
less than three particle diameters in the bulk, indicative of the
diffuse nature of the force transmission network.  However, the effect
of friction on these correlations is very weak, with the frictional
packing exhibiting only a very slight increase in local correlation.

\begin{figure}
\begin{center}
  \includegraphics[width=6cm]{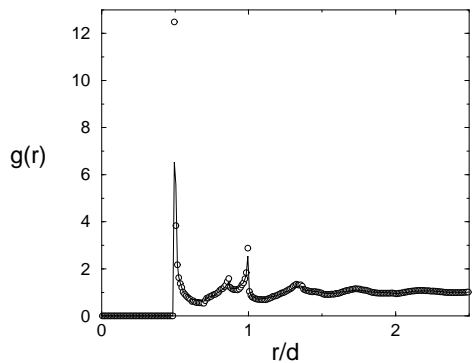} \bigskip
\caption{Radial distribution function $g(r)$ of the contact points within
  the bulk of a frictionless (circles) and frictional ($\mu=0.5$ --
  line) packing.}
\label{figure12}
\end{center}
\end{figure}

Similar to Mueth et al. \cite{mueth1}, in Fig.~\ref{figure12} we also
show the radial distribution function $g(r)$, between {\it contact
  points} inside the bulk of a frictionless ($\mu = 0$) and a
frictional packing ($\mu = 0.50$). Clearly, the frictionless packing
has a higher first peak, representative of the higher coordination of
the frictionless packing compared with the frictional one \cite{leo9},
and also local correlations between the positions of the contact
points are stronger in the case of the zero friction packing
indicating a more ordered distribution of contact points in the
system.

\subsection{Contact Geometry}
\begin{figure}[b]
\begin{center}
  \includegraphics[width=6cm]{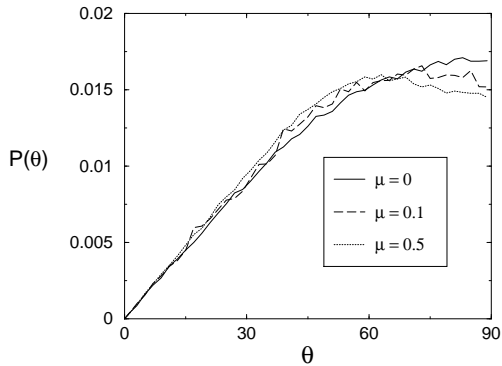}
\caption{Probability distribution functions $P(\theta)$ for particle-particle
  contact angles, where $\theta$ is defined in the local spherical
  coordinate system as the angle the particle pair makes with the
  vertical. $\theta = 0$ is a vertical contact and $\theta =
  90^{\circ}$ a horizontal contact. Packings with $\mu=0,~0.1,~0.5$
  are shown.}
\label{figure7}
\end{center}
\end{figure}

We have so far shown that computations of $P(f)$ for various particle
parameters yield essentially the same data, except for small $f$. It
is ironic then, that although the generic features of $P(f)$ are a
signature of the granularity of the system, it offers little
distinctive information on the grain-level properties of the packing.
Keeping in the spirit of particle pair information, in
Fig.~\ref{figure7} we show the probability distributions for
particle-particle contact angles defined in the local spherical
coordinate system that bonds make with respect to the vertical
(parallel to gravity direction). In Fig.~\ref{figure7} we compare
packings with different $\mu$ (=0, 0.1, 0.5) and found that the
distribution of contact angles has only a weak dependence on friction
indicating that all the systems locally appear similar. In all cases,
the majority of contact angles lie in the range
$45^{\circ}<\theta<90^{\circ}$.

\begin{figure}[b]
\begin{center}
  \includegraphics[width=6cm]{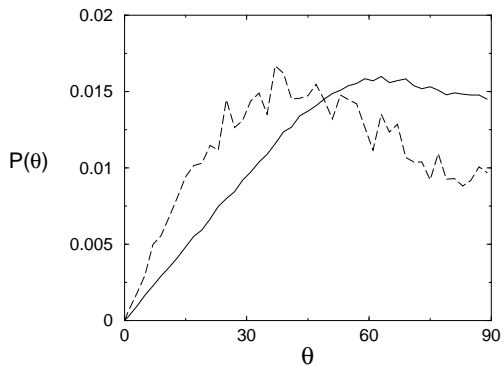}
\caption{Distribution of contact angles $P(\theta)$ of particles in contact
  for a packing with $\mu=0.5$. We distinguish between $P(\theta)$
  computed between all pairs in contact (solid line), and a sub-set of
  particle pairs whose contact force is greater than some cut-off
  threshold $f_{\text cut}$ (dashed line). Here $f_{cut} = 2{\bar f}$,
  i.e. all contacting particles whose normal contact force is greater
  than twice the average contact force. Packings for all $\mu$ exhibit
  similar behaviour.}
\label{figure8}
\end{center}
\end{figure}

It is a simple exercise to further compute $P(\theta)$ only between
particles that carry a large force, i.e. to identify or distinguish
between ``weak'' and ``strong'' forces, as Radjai and co-workers have
done for compressed systems \cite{radjai2,radjai4}. In
Fig.~\ref{figure8}, we compare $P(\theta)$ computed between all
contacting particle pairs and $P(\theta)$ computed for the subset of
particles in contact whose normal contact force $f_{n}>f_{\text cut}$,
where $f_{\text cut}$ is some given threshold value. Here we set
$f_{cut} = 2.0$, i.e. all particles whose normal contact force is
greater than the twice the average. Resolving the contact angle
distribution according to a force cut-off as in Fig.~\ref{figure8}
reveals that high force-bearing clusters are more directional and the
anisotropy grows with increasing $f_{cut}$ (not shown here).

\subsection{Contact Network}
The existence of heterogeneous force networks is supported by
experimental visualisation. Photoelastic particle packings
\cite{behringer3} or piles subject to local force perturbations
\cite{behringer5,aste1}, demonstrate inhomogeneity in the magnitude of
the forces propagating through granular assemblies - ``force chains''.
However, it is still not clear how relevant these structures are in
determining the stress state of the system. The 2D simulations of
Radjai et al.~\cite{radjai2} suggested for compressed granular
packings, a distinction can be made between the ``strong'' force
network, those particles in contact that carry a force greater than
the average normal contact force, and the ``weak'' force network, the
network of particles that experience a force smaller than the average.
In some theoretical approaches, the strong force network is assumed to
support all the stress in the system, with the weak force network
acting merely as a supporting framework to this which can essentially
be neglected \cite{claudin3}.

\begin{figure}[b]
\begin{center}
  \includegraphics[width=6cm]{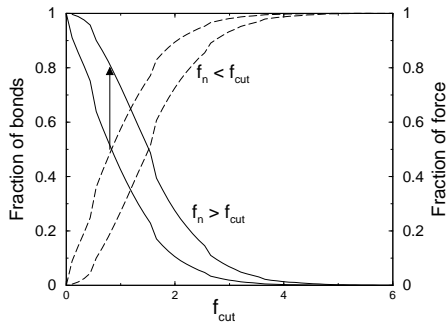}
\caption{Frictional packing ($\mu =0.5$)--contribution to the bulk average
  normal contact force and the fraction of particle contacts that make
  up this contribution, as a function of the imposed contact force
  threshold $f_{cut}$. Solid lines are the contributions from normal
  contact forces $f_n$ larger than the threshold $f_{cut}$ and dashed
  lines are for the forces that are smaller than the threshold. Thick
  solid line: fractional contribution to the average normal contact
  force for contacts with $f_n>f_{cut}$. Thin solid line: the fraction
  of contacts with $f_n>f_{cut}$. Thick dashed line: percentage
  contribution to the force coming from contacts with $f_n<f_{cut}$.
  Thin dashed line: the fraction of contacts with $f_n<f_{cut}$. The
  arrow indicates the example where $50\%$ of particle contacts
  contribute to $80\%$ of the bulk average contact force. Packings for
  all $\mu$ exhibit similar behaviour.}
\label{figure9}
\end{center}
\end{figure}

To investigate the relative importance of the force networks, we
computed the normal force that sub-networks of force chains contribute
to the bulk average contact normal force. In Fig.~\ref{figure9}, we
varied $f_{cut}$ and then computed the fraction of bonds remaining in
the force network whose contact force was greater than $f_{cut}$
(`strong' force network), and computed the contribution that this
network made to the average force.  The computation of the relative
force network contributions in Fig.~\ref{figure9} indicates only a
weak distinction between the `strong' force network for particle
contacts with $f_n\gtrsim 2$, and a weak force network with
$f_n\lesssim 2$, say. Therefore it is questionable whether the
so-called strong network actually does carry most, if not all, of the
stress in the system. For example, by going from one curve to the
other as indicated by the arrow in Fig.~\ref{figure9}, we find $50\%$
of contacts contribute approximately $80\%$ to the average contact
force.  This is a small distinction, and not nearly an order of
magnitude difference between the two networks that one might expect if
the strong forces dominated the weak phase.

\begin{figure}[b]
\begin{center}
  \includegraphics[width=2cm]{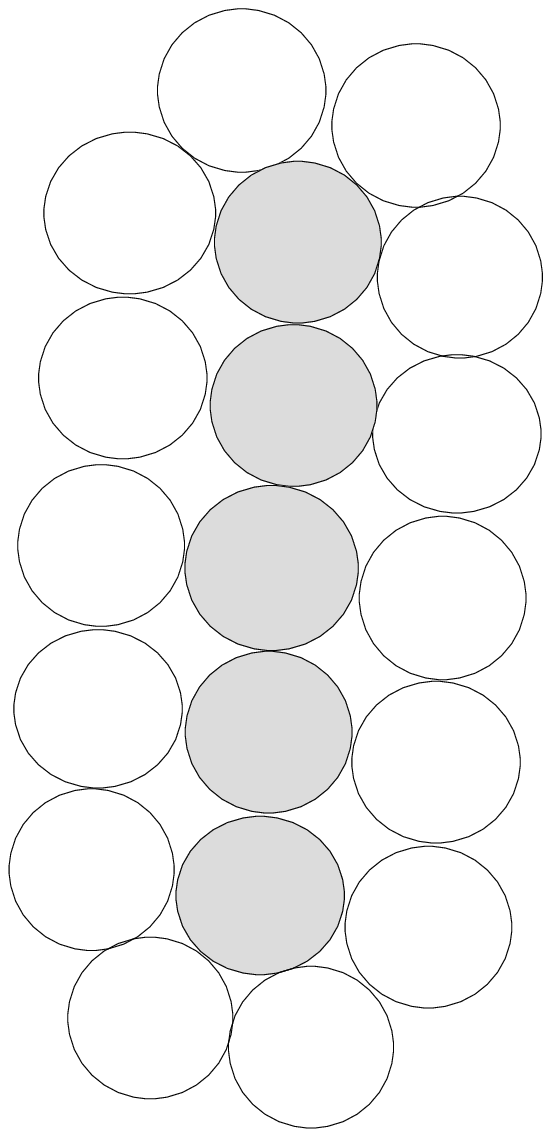} \bigskip
\caption{Schematic for computing the coordination number of a subset
  network of particles. If $f_{cut}$ determines the grey particles to
  belong to the force network, then to compute the coordination number
  of this network we need to know {\em all} contact neighbours (grey
  and white particles) of the given sub-network.}
\label{figure10}
\end{center}
\end{figure}

A related question is the stability of the relative networks. The
concept of fragility \cite{cates2} suggests in the limit one can
clearly distinguish the strong network from the weak network, the
strong phase should be a minimally coordinated, particle network. For
a 3D frictional packing this suggests a coordination number $z=4$
\cite{edwards2}. To calculate the network-averaged coordination number
of a subset of particles, the contacting neighbours of the chosen
network need to be included. In Fig.~\ref{figure10} we draw a
schematic for determining the coordination number given a sub-network
of particles (denoted by the grey particles), knowing the list of
network neighbours (white particles).

\begin{figure}[b]
\begin{center}
  \includegraphics[width=6cm]{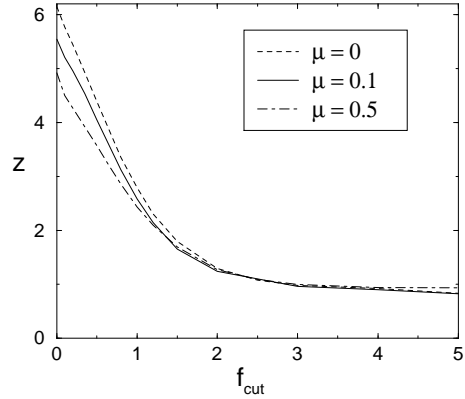}
\caption{Coordination number $z$ for packings with different $\mu$, of
  particle networks as a function of the force cut-off $f_{cut}$ that
  determines whether they belong to the network or not. }
\label{figure11}
\end{center}
\end{figure}

Computation of the coordination number $z$ for packings with different
$\mu$, over a range of cut-off values $f_{cut}$ is shown in
Fig.~\ref{figure11}. The network-averaged coordination number of
particle clusters, based on the forces that they carry, decreases
monotonically from the bulk averaged coordination $(f_{cut}=0)$ to
approximately $z=1$. It appears that $f_{cut} \approx 2$, represents
some limit in the system in the sense that for $f_{cut} > 2$, the
average size of particle clusters contributing are particle pairs,
i.e. the largest cluster that propagates large forces is only of size
two.

\begin{figure}[b]
\begin{center}
  \includegraphics[width=6cm]{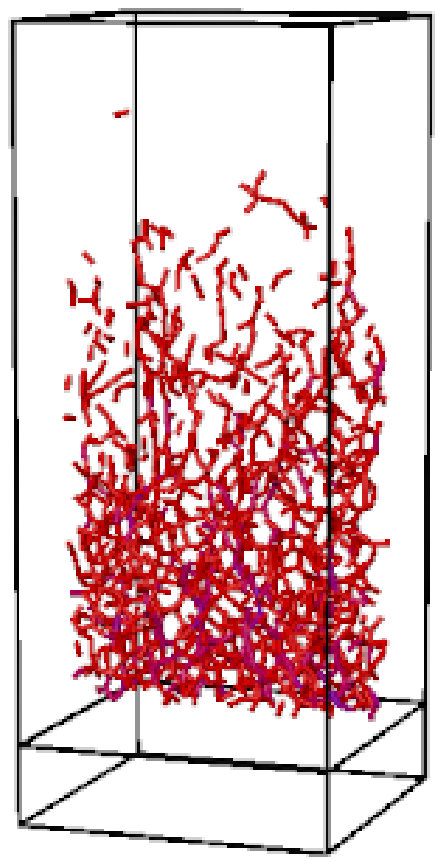}
  \includegraphics[width=6cm]{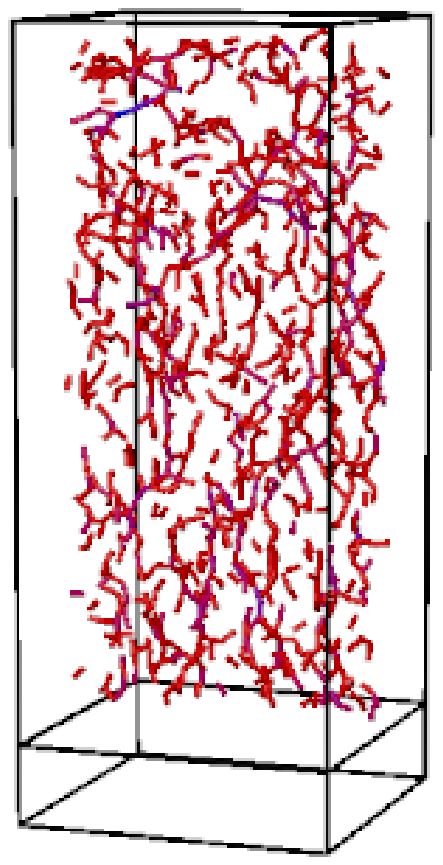}\\
\bigskip
\caption{Configurations of force networks for (left panel) the absolute forces
  with $F>2{\bar F}$, and (right) the depth-normalized forces with
  $f>2{\bar f}$. Red-blue continuous colour scheme is used to indicate
  the relative magnitude of the forces with red corresponding to
  forces closest to the lower threshold and blue are the largest
  forces.  Results for periodic packings with $\mu=0.5$ and
  $N=128000$.  Networks for different $\mu$ appear very similar. The
  black frame denotes the size of the simulation cell.}
\label{figure13}
\end{center}
\end{figure}

Additionally, we provide examples of force network realizations. In
Fig.~\ref{figure13} we show two force network configurations of a
slice approximately four particle diameters thick taken from the
centre of the large, frictional, periodic system ($N=128000$ and
$\mu=0.5$). We only show bonds whose force is greater than twice the
average. Figure \ref{figure13}(a) is the force network for the
absolute normal contact forces F without depth normalisation for
$F_{cut} = 2{\bar F}$.  This corresponds to a section through the
middle of a wide sandpile. Figure~\ref{figure13}(a) can be compared to
the 2D experimental realization in Refs.~\cite{behringer3,behringer6}.
If we show all bonds, the force network is dense with many weak
forces. This may be an indication of the relative sensitivity of the
experimental visualisation technique which clearly cannot resolve the
smallest forces. The number of large forces increases with depth
giving a clear indication of the propagation of weight down the pile.

Figure \ref{figure13}(b) is the force network for the depth-normalized
normal contact forces $f$ with $f_{cut}=2{\bar f}$. We find similar
configurations for the cylindrically confined packing. Therefore this
is the equivalent force network for a confined, frictional, unloaded
packing. Because of the weight of the particles have been renormalized
out of the force (mimicking walls that support forces), forces of all
magnitudes are seen throughout the pack. In both cases, we find that
extended force-bearing structures exist over a range of length scales,
but do not necessarily transmit the largest forces only.

\section{Conclusions}
\label{summary}
We have demonstrated that large scale simulations of granular packings
offer insight into the effects of particle friction on measurements of
the distribution of particle-particle and particle-wall contact forces
$P(f)$. Our detailed comparison between simulation and empirical fits,
obtained from experiments \cite{mueth1}, showed moderate agreement.
However, we were only able to fit our data over the full range in $f$
after renormalising our data by neglecting the smallest forces in the
system (using the partial data set). We reason that this is an
appropriate way to account for the limited resolution in experiment.
We also reiterate the fact that many simulation and numerical studies
of force distributions do not show a clear exponential tail either at
large $f$ and we believe this may partly be due to the resolution of
very small forces in such computer experiments that affect the total
normalisation parameters.

We were able to discern the influence that friction plays on $P(f)$ in
the small force region. The fraction of particle-particle contacts
that experience very small forces increases with friction even though
the total number of contacts decreases with increasing $\mu$
\cite{leo9}. Excluded volume effects rather than the functional form
of the force law appear to dominate the bulk behaviour of the system
for dense packings. Our studies of very large systems show that the
tails of $P(f)$ become marginally steeper with decreasing friction,
i.e.~$\beta$ in Eq.~\ref{equation1} increases as $\mu$ decreases. Our
ongoing work on simulating confined packings will investigate some of
these issues further \cite{landry1}.

The force-force spatial distribution function and contact point radial
distribution function indicate that spatial correlations between the
contact forces and the positions of the contact extends out only to
approximately three particle diameters. This shows that force
correlations dissipate quickly in the bulk and that the force
transmission network propagates locally but becomes diffuse rapidly.
On introducing a force cut-off scheme to analyse force-bearing
structures, we found no clear evidence for distinct `weak' and
`strong' force phases.

In general this discussion was only concerned with unloaded or
unperturbed amorphous granular packings. In this sense we have
provided information on the ``reference state'' of a granular material
from a microstructural point of view. This state is rather insensitive
to friction and is primarily determined by construction history
\cite{cates9}. Although we have not investigated the perturbed state
or response function of these systems \cite{behringer5}, it is likely
that particle properties play a much more significant role in the
response of a granular system than in the static state. Some
theoretical treatments \cite{claudin3,claudin1} on force chain
analysis may benefit from the information of this unperturbed system
when calculating the resulting response of such a system under some
force perturbation. In fact, the contact angle distribution in
Fig.~\ref{figure7} seems to suggest that the `splitting angle'
$\theta_{s}=90^{\circ}-\theta$, in the language of
Ref.~\cite{claudin1}, does seem lie predominantly in the range
$0<\theta_{s}<30^{\circ}$. Comparing the experimental visualisation in
Ref.~\cite{behringer5} and the theoretical model in
Ref.~\cite{claudin1}, the force chain analysis can be thought of as a
superposition of force chains on top of the background force network.

This work was supported by the Division of Materials Science and
Engineering, Office of Science, U.S. Department of Energy. This
collaboration was performed under the auspices of the DOE Center of
Excellence for the Synthesis and Processing of Advanced Materials.
Sandia is a multiprogram laboratory operated by Sandia Corporation, a
Lockheed Martin Company, for the United States Department of Energy
under Contract DE-AC04-94AL85000.


\begin{thebibliography}{10}

\bibitem{nagel1}
H.~M. Jaeger and S.~R. Nagel,  Science {\bf 255,} 1523 (1992).

\bibitem{combo1}
A.~S. Clarke and H. Jonsson, Phys. Rev. E {\bf 47,} 3975 (1993); R.
  Jullien, P. Jund, and D. Caprion, Phys. Rev. E {\bf 54,} 6035
  (1996); S. Torquato and F.~H. Stillinger, J. Phys. Chem. B {\bf
    105,} 11849 (2001).

\bibitem{leo9}
L.~E. Silbert, D. Erta{\c s}, G.~S. Grest, T.~C. Halsey, and D. Levine,  Phys.
  Rev. E {\bf 65,} 031304 (2002).

\bibitem{mueth1}
D.~M. Mueth, H.~M. Jaeger, and S.~R. Nagel,  Phys. Rev. E {\bf 57,} 3164
  (1998).

\bibitem{coppersmith1}
S.~N. Coppersmith, C. h.~Liu, S. Majumdar, O. Narayan, and T.~A. Witten,  Phys.
  Rev. E {\bf 53,} 4673 (1996).

\bibitem{nagel5}
D.~L. Blair, N.~W. Mueggenburg, A.~H. Marshall, H.~M. Jaeger, and S.~R. Nagel,
  Phys. Rev. E {\bf 63,} 041304 (2001).

\bibitem{nagel6}
N.~W. Mueggenburg, H.~M. Jaeger, and S.~R. Nagel, cond-mat/0204533
  (unpublished).

\bibitem{lovoll1}
G. L$\phi$voll, K.~J. M{\.a}l$\phi$y, and E.~G. Flekk$\phi$y,  Phys. Rev. E
  {\bf 60,} 5872 (1999).

\bibitem{charmet1}
O. Tsoungui, D. Vallet, and J.-C. Charmet,  Gran. Matter {\bf 1,} 65 (1998).

\bibitem{radjai1}
F. Radjai, M. Jean, J.-J. Moreau, and S. Roux,  Phys. Rev. Lett. {\bf 77,} 274
  (1996).

\bibitem{radjai2}
F. Radjai, D.~E. Wolf, M. Jean, and J.-J. Moreau,  Phys. Rev. Lett. {\bf 80,}
  61 (1998).

\bibitem{leo7}
L.~E. Silbert, D. Erta{\c s}, G.~S. Grest, T.~C. Halsey, D. Levine, and S.~J.
  Plimpton,  Phys. Rev. E {\bf 64,} 051302 (2001).

\bibitem{cundall1}
P.~A. Cundall and O.~D.~L. Strack,  G$\acute{e}$otechnique {\bf 29,} 47 (1979).

\bibitem{walton2}
O.~R. Walton,  Mech. Mat. {\bf 16,} 239 (1993).

\bibitem{wolf1}
J. Schafer, S. Dippel, and D.~E. Wolf,  J. Phys. I France {\bf 6,} 5 (1996).

\bibitem{landry1}
J.~W. Landry, L.~E. Silbert, G.~S. Grest, and S.~J. Plimpton, in preparation
  (unpublished).

\bibitem{radjai3}
F. Radjai, S. Roux, and J.~J. Moreau,  Chaos {\bf 9,} 544 (1999).

\bibitem{thornton1}
C. Thornton,  Kona Powder and Particle {\bf 15,} 81 (1997).

\bibitem{snoeijer1}
J.~H. Snoeijer, M. van Hecke, E. Somfai, and W. van Saarloos, cond-mat/0204277
  (unpublished).

\bibitem{ohern2}
C.~S. O'Hern, S.~A. Langer, A.~J. Liu, and S.~R. Nagel,  Phys. Rev. Lett. {\bf
  88,} 075507 (2002).

\bibitem{coppersmith2}
C. h.~Liu, S.~R. Nagel, D.~A. Schecter, S.~N. Coppersmith, S. Majumdar, O.
  Narayan, and T.~A. Witten,  Science {\bf 269,} 513 (1995).

\bibitem{radjai4}
F. Radjai and D.~E. Wolf,  Gran. Matter {\bf 1,} 3 (1998).

\bibitem{behringer3}
L. Vanel, D. Howell, D. Clark, R.~P. Behringer, and E. Clement,  Phys. Rev. E
  {\bf 60,} R5040 (1999).

\bibitem{behringer5}
J.~G. Geng, D.~W. Howell, E. Longhi, R.~P. Behringer, G. Reydellet, L. Vanel,
  and E. Clement,  Phys. Rev. Lett. {\bf 87,} 035506 (2001).

\bibitem{aste1}
T. Aste, T. DiMatteo, and E.~G. d'Agliano,  J. Phys.:Condens. Matter {\bf 14,}
  2391 (2002).

\bibitem{claudin3}
J.~E.~S. Socolar, D.~G. Schaeffer, and P. Claudin,  Eur. Phys. J. E {\bf 7,}
  353 (2002).

\bibitem{cates2}
M.~E. Cates, J.~P. Wittmer, J.-P. Bouchaud, and P. Claudin,  Phys. Rev. Lett.
  {\bf 81,} 1841 (1998).

\bibitem{edwards2}
S.~F. Edwards,  Physica A {\bf 249,} 226--231 (1998).

\bibitem{behringer6}
J.~G. Geng, E. Longhi, R.~P. Behringer, and D.~W. Howell,  Phys. Rev. E {\bf
  64,} 060301 (2001).

\bibitem{cates9}
J.~P. Wittmer, M.~E. Cates, and P. Claudin,  J. Phys. I France {\bf 7,} 39
  (1997).

\bibitem{claudin1}
J.-P. Bouchaud, P. Claudin, D. Levine, and M. Otto,  Eur. Phys. J. E {\bf 4,}
  451 (2001).

\end{thebibliography}
\end{document}